\documentclass[5p,sort,compress]{elsarticle}
\usepackage{amsfonts}
\usepackage{amsmath}
\usepackage[labelfont=bf,justification=raggedright,singlelinecheck=false]{caption}
\usepackage{array}
\usepackage{tabu}
\usepackage{multirow}
\usepackage{graphicx}
\usepackage{amsmath}
\usepackage{algpseudocode}
\usepackage{tikz}
\usepackage[linesnumbered,ruled,vlined]{algorithm2e}
\usepackage{pdflscape}
\usepackage{rotating}
\usepackage{wrapfig}
\usepackage[inline, shortlabels]{enumitem}
\usepackage{mathtools}
\usepackage{amsmath}

\newcolumntype{v}{>{\centering\arraybackslash}m{.18\linewidth} }
\begin{document}
\journal{arxiv}
\title{High Capacity Image Data Hiding of Scanned Text Documents Using Improved Quadtree}
\author[fum]{Seyyed Hossein Soleymani}
\ead{seyyedhosein.soleymani@mail.um.ac.ir}
\author[fum]{Amir Hossein Taherinia\corref{cor1}}
\ead{taherinia@um.ac.ir}
\cortext[cor1]{Corresponding author}
\address[fum]{Computer Engineering Department, Ferdowsi University of Mashhad, Mashhad, Iran }
\begin{abstract}
In this paper, an effective method was introduced to steganography of text document in the host image. In the available steganography methods, the message has a random form. Therefore, the embedding capacity is generally low. In the proposed method, the main underlying idea was the sparse property of scanned documents. The scanned documents were converted from gray-level form to binary values by halftoning idea and then the information-included parts were extracted using the improved quadtree and separated from document context. Next, in order to compress the extracted parts, an algorithm was proposed based on reading the binary string bits, ignoring the zero behind the number, and converting them to decimal values. Embedding capacity of the proposed method is higher than that of other available methods with a random-based message. Therefore, the proposed method can be used in the secure and intangible transfer of text documents in the host image. 
\end{abstract}
\maketitle
\section{Introduction}
In recent years, data hiding has been widely used in different fields. According to the different application of data hiding in today's life, this science has been divided into different subcategories based on its application. Two main subcategories of data hiding include steganography and watermarking \cite{ref001}. The important similarity among these subcategories is hidden messages, watermarks, or information. This similarity results in same sense in various application of both subcategories.

In watermarking, some bite in robust, fragile, or semi-fragile form, is embedded as a watermark in an image to protect the image against claiming intellectual property, forging, tampering, etc. \cite{ref002,ref003,ref004,ref005}. Robust watermarking is the ability to retrieve the watermark after intentional attacks, such as image editing and manipulation, and unintentional attacks, such as compression and image quality enhancement processes. Besides, fragile and semi-fragile watermarking is used to determine the manipulation of the image and retrieve manipulation parts. However, the main aim of steganography is to hide information in a form of a message in an image, but the message and its secure and imperceptible transmission are of central importance. In fact, in steganography, in the most cases the image itself does not matter, transferring a larger and more imperceptible volume of the message is important. There is a challenge among the criteria of robustness, capacity, and invisibility. Accordingly, increasing each one can reduce the number of other criteria~\cite{ref004}. Therefore, one of the key points in data hiding is to design an approach by considering the main goal and preserve the other goals. 

In steganography methods, embedding the message information in an image is carried out in spatial and frequency domains. Generally, embedding in spatial domain is performed in less important bits of the image without considering special conditions, which it is possible to guess whether it is covered or not by evaluating the image histogram. In order to increase the power of LSB-based methods, LSB++ method is proposed, which aims to avoid embedding in certain pixels of the image in order to maintain the image histogram~\cite{ref006}. Some steganography methods, which are known as reversible methods, try to achieve relatively complete retrieval of host image in the receiver after message extraction~\cite{ref007,ref008,ref009,ref010}. In terms of classification, the methods with higher capacity of steganography including histogram transmission-based methods~\cite{ref011,ref012}, Quantization-based methods~\cite{ref013}, interpolation-based methods, Pixel Value Difference (PVD) based method~\cite{ref5} and other methods~\cite{ref3}. All available steganography methods assume that the message has a random form, which reduces the embedding capacity of the proposed algorithm. 

According to the digitalization of academic and official documents and letters, the secure and imperceptible document image transferring and storing as a message is needed. Although in steganography methods, several attempts have been made to send any message as a random bit, there has not been many efforts to embed and send scanned documents and administrative letters as a message in a safer way~\cite{ref8}. In the present studies, a high-capacity steganography for scanned documents as a message in an image was proposed to embed a significant amount of information in a host image. 
 The paper proceeds as follows. In section~\ref{lbl.relatedworks}, the related methods are introduced. In section~\ref{lbl.proposedmethod}, the proposed method and its sub-steps are discussed. Finally, the results and conclusion are presented in section~\ref{lbl.experimentalresult} and~\ref{lbl.conclusions}, respectively.

\section{Related Works}\label{lbl.relatedworks}
In this section, the strongest methods and most related methods were respectively evaluated on the embedding capacity of the random message and nonrandom message. Besides, the main ideas along with embedding capacity were briefly discussed.  

In~\cite{ref1}, AMBTC compression-based steganography in an image was presented, in which the input image was converted to $\frac{1}{4}$ of its original size, and then the compressed pixels were dispersed in a particular way in an image, which was the same size as the original image. The empty pixels were calculated through the proposed interpolation ASAI method. An AMBTC compression resulted in a reduction in quality of the embedded image. In this method, amount of PSNR and embedding capacity was 33.39 dB and 1.2 bpp that both criteria had low values. 

In~\cite{ref2}, the main image was converted to a larger image using interpolation, which is called cover image. The difference between the interpolated pixels and the maximum pixels of a neighbor in blocks was used to determine the amount of bit that could be embedded in an interpolated pixel. Therefore, the visual quality of the image was further preserved. The quality of the embedded image was 33.85 dB in terms of PSNR and its embedding capacity was 1.79 bpp.

In~\cite{ref3}, a method was proposed to have fewer changes in the image and used the index function to calculate the basic pixels. The pixels of each block were being displaced based on the calculated amount for basic pixels. Amount of embedded bit for each pixel was calculated by the difference between each pixel in a block with a base zero number pixel. In this method, in order to maintain the image quality in embedding phase, the ranges table was used and the gray-level of pixels was divided into groups. The embedded image quality was 28.45 dB and its embedding capacity was 2.45 bpp. Although the value of embedding capacity was relatively enough, the embedded image quality was fewer than the embedded image with over 35 dB.

In~\cite{ref4}, weighted matrix and modular sum resulting from the multiplication of a weighted matrix and image pixels was used to calculate the location of the embedded image and the message information was embedded in three less important bits of the image. In this method, a different weighted matrix was used for different blocks, which resulted in an increase in security of steganography. In terms of PSNR, the quality of the embedded image was 37.97 dB and storage capacity was 2.97 bpp.

In~\cite{ref5}, an Octanary PVD-based high-capacity steganography method was proposed. It means in each 3$\times$3 block, all neighbor pixels of central pixel were paired with central pixels and then the amount of embedded bit in that pair, as well as whether the pixels were an edge or not, was decided. In this method, as in many similar methods, a ranges table was used to determine the interval of each pixel placement. After embedding, a phase, under the name of the readjustment, was used on embedded pixels to align on gray level correctly and not go beyond the permissible range. In terms of PSNR metric, the quality of the embedded image was 38.98 dB and storage capacity was about 3.72 bpp, which both criteria had high value. 

In~\cite{ref6}, a high capacity steganography was proposed that used a genetic algorithm to choose the suitable embedding order in a 4$\times$4 block. In fact, this paper indicated that conventional embedding order did not have the best image quality and embedding capacity. This method aimed to convert the best embedding order to an optimization problem and solve it using a genetic algorithm. Besides, this method determined the best LSB bits to have higher capacity. The embedding capacity was 3.95 bpp and the quality of embedded image was about 35 dB. 

In~\cite{ref7}, a high capacity steganography method based on determining the image edges was proposed. In this paper, in order to determine the edge, three important bits of each pixel were used in different conventional edge detection methods including Sobel, Canny and fuzzy based methods. Then image pixels were divided into the edge and non-edge pixels. In each edge pixel and non-edge pixel, 4 and 2 bits of the message were respectively embedded. The embedding capacity and quality was 4.7 bpp and 26 dB, respectively. As seen, this method has high embedding capacity but its image quality is fewer than similar method. 

In~\cite{ref8}, a steganography method was proposed, which used the scanned document as a message. This method aimed to perform an algorithm to compress the message by considering the sparse property of the message. Two important achievements of this method were to convert gray-level image to binary image using halftone method and convert the binary string to their equivalent decimal value. One of the main disadvantages of this method was to code the background area in addition to areas containing content. Embedding capacity and quality of embedded image were 5.25 bpp and 36 dB, respectively.

By evaluating the recent high capacity steganography methods it can be found that the most of these methods did not consider the type of input message, i.e., the message was considered as random. In scanned text documents and letters, the background is white and the content part is black, therefore, if input message is produced by scanned text documents, official letters, and etc. the scanned document could be converted to a sparse message through a special method. By considering the message as a sparse form, the high capacity steganography could be presented to hide the message with large size.

 \begin{figure*}
   \centering
   \includegraphics[width=1\textwidth]{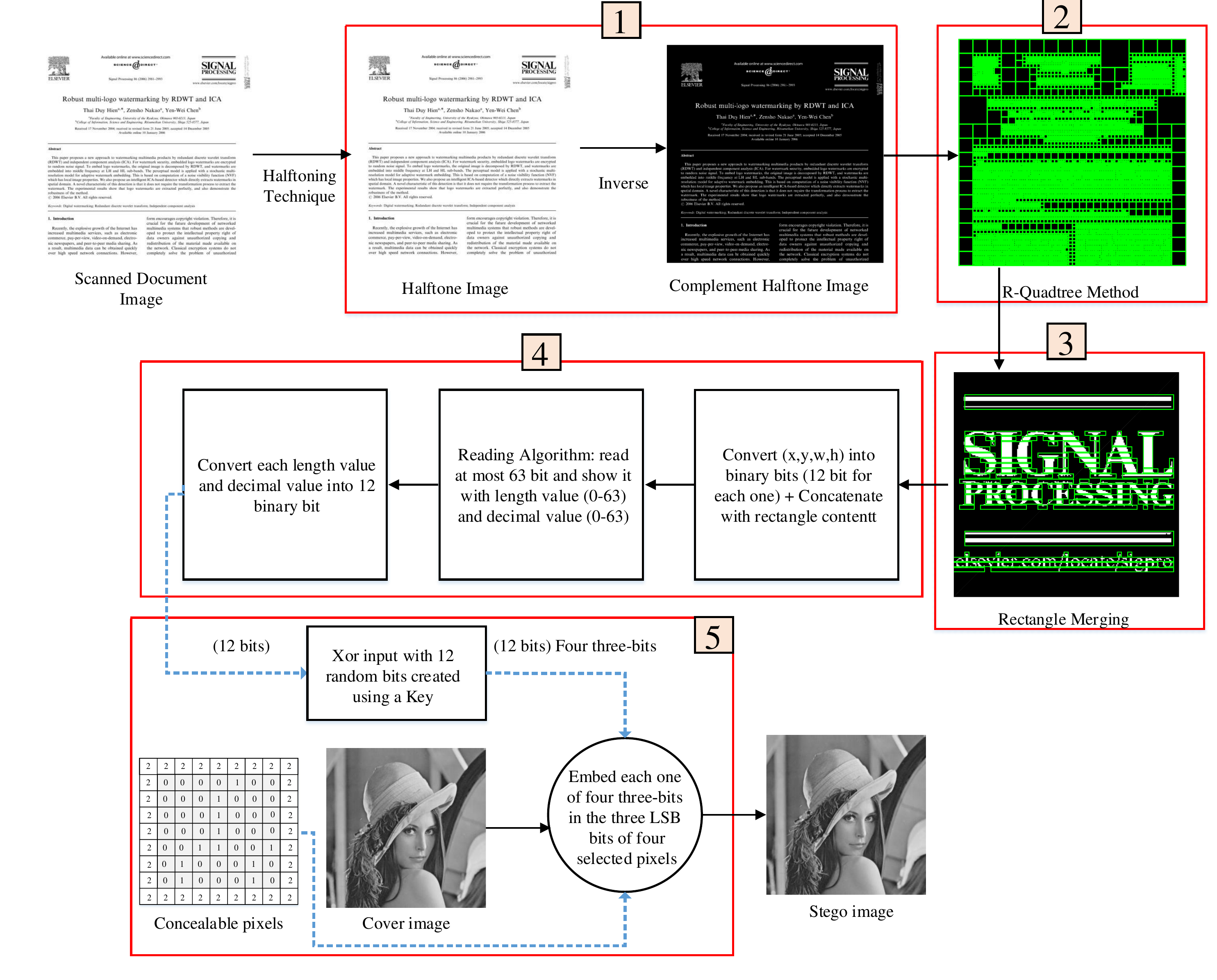}
   \caption{Diagram of document embedding algorithm in the host image.}\label{fig.Embedflowchart}
 \end{figure*} 
 
   \begin{figure*}
     \centering
     \includegraphics[width=1\textwidth]{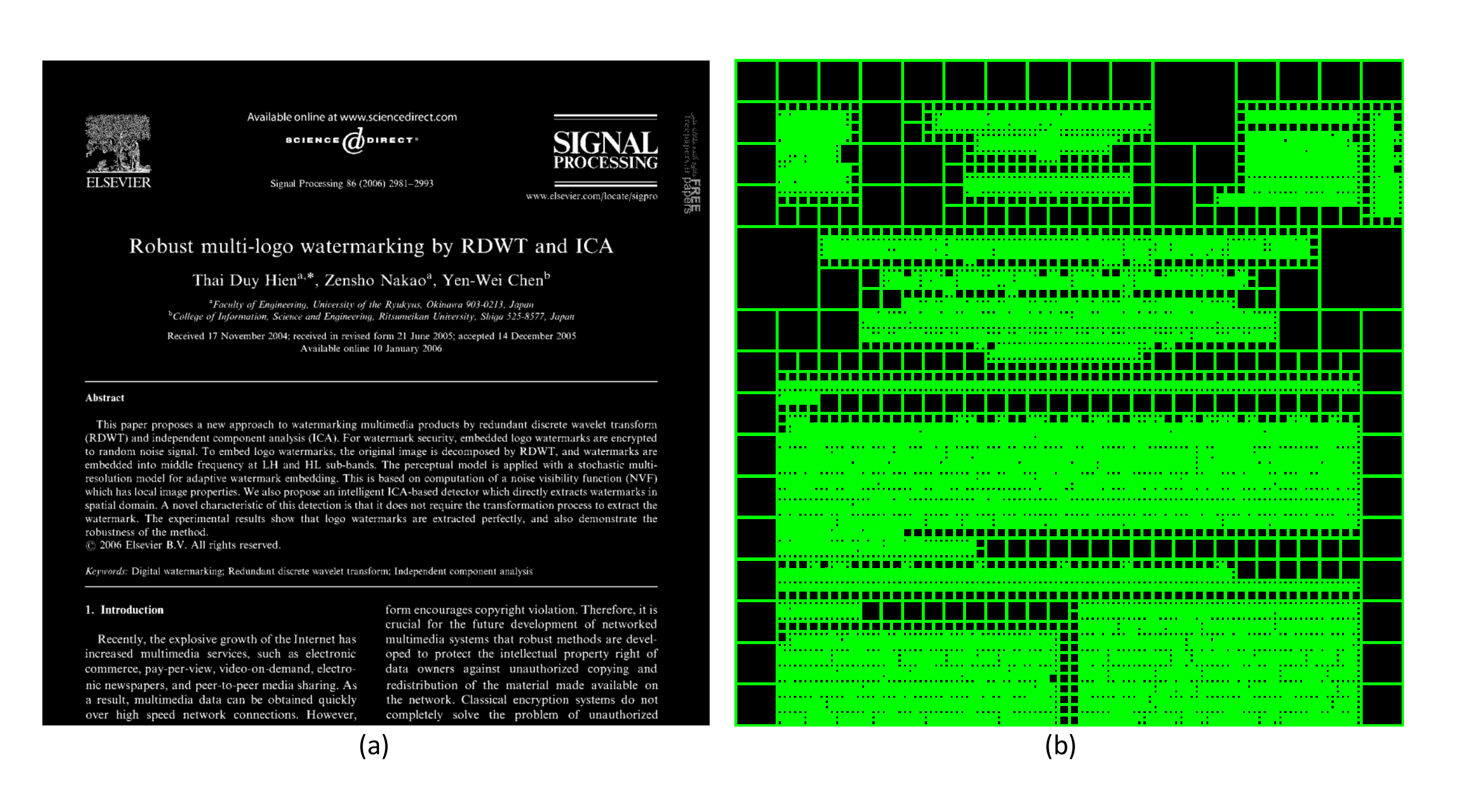}
     \caption{a: complement form of the halftone image, b: output of R-Quadtree algorithm.}\label{fig.main_decomposition}
   \end{figure*} 
   
   	\begin{figure*}
   	  \centering
   	  \includegraphics[width=1\textwidth]{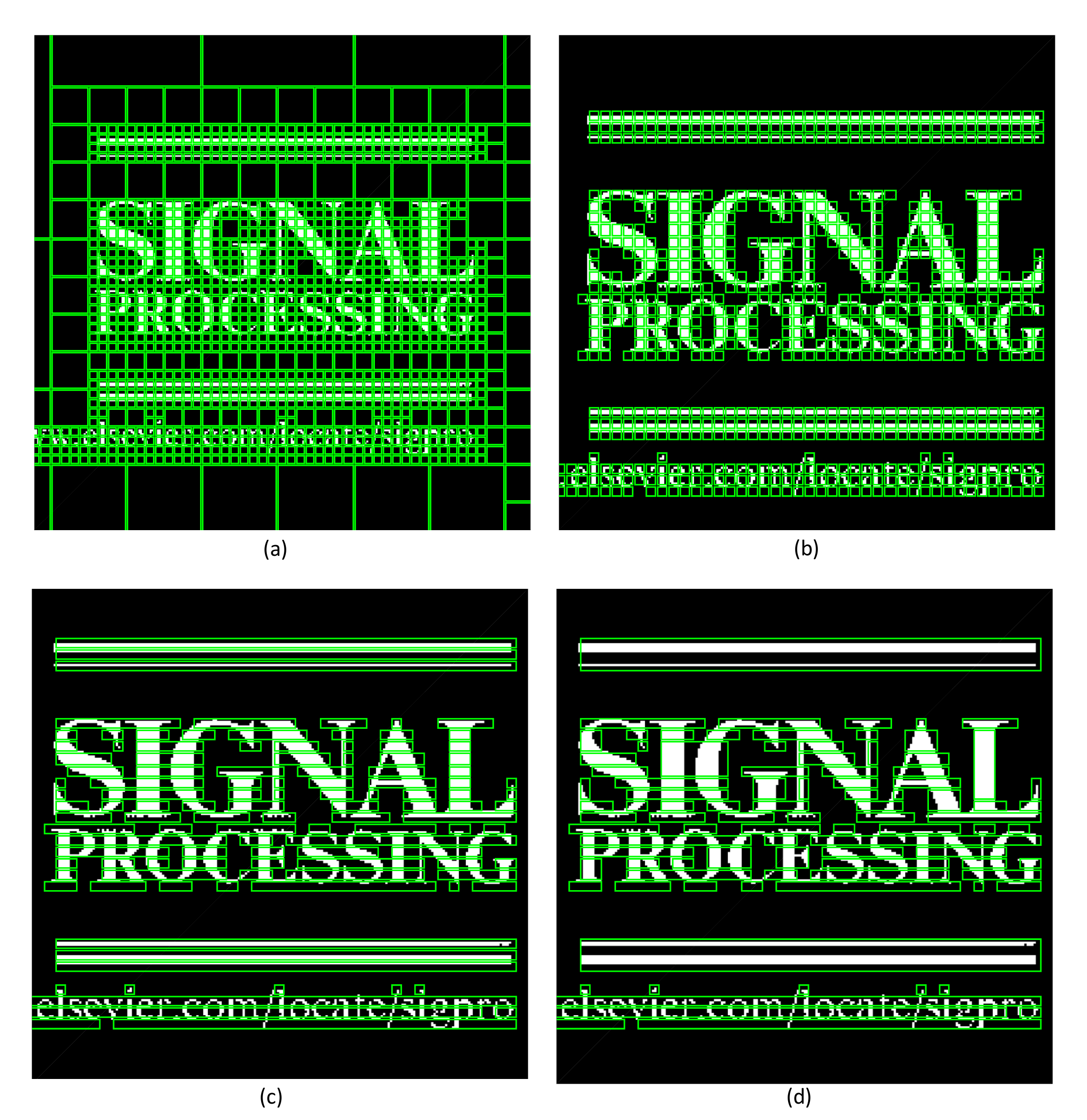}
   	  \caption{a: output of R-Quadtree algorithm; b: presentation of content-contained rectangles; c: horizontal integration of rectangles; d: vertical integration of rectangles.}\label{fig.four_Decomposition}
   	\end{figure*} 

\section{Proposed Method} \label{lbl.proposedmethod}

In Fig.~\ref{fig.Embedflowchart} the proposed algorithm is presented. As seen in Fig.~\ref{fig.Embedflowchart}, the main operations are shown with 5 numbers. In the first step, which is called pre-processing stage, the halftone image is calculated by scanned document and the sparse message is obtained. In the second step, the operation of content separation from the background is performed using the improved quadtree method. Then in the third step, in order to reduce the number of blocks, the blocks are combined in two vertical and horizontal directions, which results in a reduction in the need of maintaining coordinates for blocks. In next step, in order to achieve more compression, the decimal coding is conducted on content. Finally, in the fifth step, the embedding of compressed message has been shown
 in three low importance pixels of host image. In extraction phase, information is extracted from LSB pixels and by considering the coordinates of each block, its content is placed in that block. The details are given as follows.

\subsection{Preprocessing step}

In this stage, the gray-level scanned document by halftone method is converted to embeddable bits. In fact, halftone method creates a binary image from the gray-level of the scanned text image and reduces its size 8 times. It means each 8-bit pixel is shown by only one bit. Therefore, it is possible to embed the scanned document in host image, but the number of bits in scanned documents is very high that make it impossible by conventional steganography. There are various methods to calculate the halftone image from the document image, which are divided into two group including the error-diffusion based~\cite{ref10,ref11,ref12} and dither based methods~\cite{ref13,ref14}. Both halftone methods, by considering the correlation between proximate pixels, decide to convert each pixel (ranged from 0 to 255) to 0 or 1. In the present study, the proposed method of~\cite{ref10} was used to calculate the halftone image. In order to convert the halftone image to approximate image of the original gray-level image, there are different inverse halftone calculation methods, which aim to retrieve the original image more realistic~\cite{ref15,ref16}. The reverse halftone image calculation is converted each 0 and 1 bit to an integer value between 0 and 255. The simplest method to calculate the inverse halftone image is to apply a Gaussian-softening filter on halftone image. Therefore, the relatively good gray-level image is obtained. In the present study, the mentioned method was employed to calculate the reverse halftone image. The advantage of halftone method to simple thresholding on text document is to act smartly when the image document containing a contentious spectrum of pixels such as a personal image. In this case, the simple thresholding is not able to estimate the pixels values from zero or one bits.

\subsection{Embedding step}

The embedding algorithm contains four sub-steps. In the first sub-step, an algorithm was proposed to separate the parts containing text, image, signature, or any signs in the document using improved quadtree method. In the second sub-step, in order to reduce the number of rectangles, their merge was proposed. In this step, more compression in the message was achieved, while there was more ability for compressing. Therefore, in third sub-steps, an idea was used to convert the binary message to decimal ones and ignore the zero bits behind the binary bits string. Finally, in the fourth step, the compressed bit strings were embedded in three low importance bits of the pixel with a certain structure.

\subsubsection{Improved quadtree}
The halftone image of a scanned document with the binary display can include text, image, table, and any other signs, which are shown with zero bits and placed on a surface of a white background with one bits. Definitely, the coding and maintenance of bits of a background are not worth and it is best to separate the document content from the background and only the document content is embedded in the host image. Accordingly, an improved quadtree method is proposed. The main method of quadtree~\cite{ref17} is performed on the gray-level of the image with square dimensions and power of 2. In the present study, enhanced quadtree idea was proposed, which had the ability to be applied to any image dimensions. The pseudo code of this method was shown in Alg.~\ref{algo.rqtree}. As seen, the input of this algorithm is a gray-level image of the scanned document, which can be rectangular. Besides, it has two thresholds of $T1$ and $T2$ and its outputs are set to $x$, $y$, $w$, and $h$ values related to rectangles without the ability to divide into more sub-rectangles. In this set, $x$ and $y$ are left and upper side coordinates and $w$ and $h$ are width and height of the rectangle.
\begin{algorithm}[h]
 \KwData{$HalftonImg$, $MinimumLength$, $Treshold$}
 \KwResult{ $AllBlocks(x, y, w, h)$ // Each row contain $(x, y, w, h)$ coordinate of a non-decomposable rectangle. }
 $(m, n)=size(HalftonImg)$\;
 $AllBlocks(1,1:4)=(1, 1, m, n)$\;
 $Continue_{Flag}=$True\;
 \While{There is exist a block $B$ where $min(B_{w}, B_{h}) > 2\times MinimumLength$ And $(max_{grayval}(B)-min_{grayval}(B))>=Treshold)$}{
   $ToM=\lfloor B_{w}/2 \rfloor$\;
   $ToN=\lfloor B_{w}/2 \rfloor$\;
   $B_{1}=B(B_{x}, B_{y}, B_{w}, B_{h})$\;
   $B_{2}=B(B_{x}, B_{y}+ToN, B_{w}, B_{h}-ToN)$\;
   $B_{3}=B(B_{x}+ToM, B_{y}, B_{w}-ToN, B_{h})$\;
   $B_{4}=B(B_{x}+ToM, B_{y}+ToN, B_{w}-ToM, B_{h}-ToN)$\;
   Delete $B$ from $AllBlocks$\;
   Add $B_{1}$, $B_{2}$, $B_{3}$ and $B_{4}$ to AllBlocks\;  
 }
 Return $AllBlocks$;
 \caption{Rectangular-based Quadtree (R-Quadtree)}
      \label{algo.rqtree}
\end{algorithm}

In fact, first, the algorithm considers the input image as a rectangle and if the difference between its largest and smallest pixel is more equal than the threshold $T1$ and also the minimum width and height of the rectangle is 2 times larger than the threshold $T2$, it means that the rectangle should be divided into sub-rectangles. Therefore, the algorithm has divided the rectangle into four sub-rectangles, which it is possible to be in a different size (different in one row and column). This process is performed repeatedly on all sub-rectangles and it would stop when there is no other sub-rectangle with division conditions. The threshold $T1$ for the halftone image is one because of its binary property. Fig.~\ref{fig.main_decomposition}a~and~\ref{fig.main_decomposition}b show the square parts of halftone image and R-Quadtree algorithm output, respectively. Most of these rectangles contain no special information or text, therefore it is not necessary to keep their content and coordinates.

\subsubsection{Proximate Rectangles Integration}
As seen in Fig.~\ref{fig.Embedflowchart}-step(2), not all rectangles contain information; therefore, it is only needed to keep the content and coordinates of rectangles that contains information. On the other side, the number of these rectangles can be high and maintain 4 numbers as coordinates for each rectangle increase in the size of the final message and reduce the advantage of its compression. Therefore, the integration of proximate rectangles is proposed. The integration algorithm is shown in~Alg.\ref{algo.rectmerg}. Input of algorithm is $FAllBlocks$, all rectangles that contain information, and its output is integrated rectangles in both vertical and horizontal directions. This algorithm attempts to merge rectangles by scanning neighbor rectangles horizontally and then vertically. The output of proximate rectangles integration is shown in~Fig.~\ref{fig.four_Decomposition}. Fig.~\ref{fig.four_Decomposition}b shows all rectangles that contain important information. As shown in~Fig.~\ref{fig.four_Decomposition}c and~Fig.~\ref{fig.four_Decomposition}d, in the first scan, all rectangles that are neighbors horizontally and have horizontal integration condition are integrated and in the second scan, all output rectangles of the first scan, which have vertical integration condition, are integrated.

\begin{algorithm}[h]
 \KwData{$InterestBlocks$}// InterestBlocks contains all foreground blocks.\; 
 \KwResult{ $MergedBlocks$ }
 $Counter=0$\;
 \While{Count($InterestBlocks$)}{
   $(x,y,w,h)=InterestBlocks(1,1:4)$\;
   Find $B$ where $x==B_{x}$ And $y+h=B_{y}$ And $w=B_{w}$\;
   \If{Such $B$ is exists}{
      $h=h+B_{h}$\;
      $InterestBlocks(1,1:4)=(x,y,w,h)$\; 
      Delete $B$ from  $InterestBlocks$ \;
      }
      \Else
      {
      $Counter=Counter+1$\;
       $MergedBlocks(Counter,1:4)=InterestBlocks(1,1:4)$\;
	   $InterestBlocks=InterestBlocks(2:end,1:4)$\;
      }
 }
$Counter=0$\;
$InterestBlocks$=$MergedBlocks$\;

 \While{Count($InterestBlocks$)}{
   $(x,y,w,h)=InterestBlocks(1,1:4)$\;
   Find $B$ where $y==B_{y}$ And $x+w=B_{x}$ And $y=B_{y}$\;
   \If{Such $B$ is exists}{
      $w=w+B_{w}$\;
      $InterestBlocks(1,1:4)=(x,y,w,h)$\; 
      Delete $B$ from  $InterestBlocks$ \;
      }
      \Else
      {
      $Counter=Counter+1$\;
       $MergedBlocks(Counter,1:4)=InterestBlocks(1,1:4)$\;
	   $InterestBlocks=InterestBlocks(2:end,1:4)$\;
      }
 }

 Return $MergedBlocks$;
 \caption{Rectangle Merging}
      \label{algo.rectmerg}
\end{algorithm}

\subsubsection{Decimal Coding}
By ignoring the zeros in the left side of a binary bit string, decimal coding algorithm aims to read longer binary bit strings and convert them to their equivalent decimal value. In the first step, in order to increase the number of zeros in the message, the bits of the original halftone image is complemented (This work is done in step 1 of Fig.~\ref{fig.Embedflowchart}). Then, the information of document dimension (two 20-bits number), coordinates of all blocks (four 12-bits number for coordinates of each block, width and length), and its contents are attached in form of a vector and the decimal coding algorithm is employed to read the binary bit string and convert them to decimal values. In decimal coding algorithm (Alg.~\ref{algo.reading}), first, a string with a maximum length of 63 bits is read such that its decimal value be less or equal to 63. The numerical value of string length and its decimal value are respectively named $length_{val}$ and $dec_{val}$. Then, each one of the $length_{val}$ (an integer value in range 0 to 63) and $dec_{val}$ (an integer value in range 0 to 63) are converted to 6-bit binary number and it is used in final embedment sub-step.

\begin{algorithm}[h]
 \KwData{Vector message bits}
 \KwResult{ $dec_{val} $, $length_{val}$  }
 $Length_{Counter}=5$\;
 $DeciVal_{Prev}=0$\;
 $Continue_{Flag}=$True\;
 \While{$Continue_{Flag}==$ \textrm{True}}{
  $Selected_{Msg}=$ReadFromMessage$(1:Length_{Counter})$\;
  $Temp_{DeciVal}=$BinToDeci$(Selected_{Msg})$\;
  \If{$Temp_{DeciVal}\geq 64$}{
   $dec_{val}=DeciVal_{Prev}$\;
   $length_{val}=Length_{Counter}-1$\; 
   $Continue_{Flag}=$False\;
   }
  \If{($(Continue_{Flag}==$True$)$ And $(Length_{Counter}==64)$)}{
$dec_{val}=Temp_{DeciVal}$\; 
$length_{val}=Length_{Counter}-1$\; 
$Continue_{Flag}=$False\; 
   }
   $Length_{Counter}=Length_{Counter}+1$\;
   $DeciVal_{Prev}=Temp_{DeciVal}$\; 
   
 }
 \caption{Reading algorithm of each bit stream}
      \label{algo.reading}
\end{algorithm}

\subsubsection{Final Embedding }
At this point, the information bit strings are compressed using halftone, R-Quadtree, and decimal coding ideas. In order to embed the image and increase the invisibility property, the pixels in the more complex texture of image are considered as embeddable and pixels in smooth texture of image are considered non-embeddable. For calculating the embeddable and non-embeddable pixels, first, three fewer important bits of all pixels became zero and then in 3$\times$3 proximity, standard deviation (SD) is calculated. If the SD is more than threshold $T3$ the central pixel is embeddable otherwise, it is non-embeddable. The advantage of zeroing the three less important bits of all pixels in host image before calculating SD is to obtain same inputs for calculating embeddable and non-embeddable pixels in both embedding and extraction phases and it is possible to use the three fewer important bits of the embeddable pixel to hide the message. The calculation of SD in 3$\times$3 proximity is shown in Eq.~\ref{Eq.stdDev}.

\begin{equation}\label{Eq.stdDev}
\centering
 std_{kernel} =\frac{1}{(9-1)}\sum_{x=1}^{3}\sum_{y=1}^{3}\left (
\hat{f}_{kernel}(x_{i,j})- \bar{x}_{kernel} \right )^2.
\end{equation}
\begin{equation}\label{Eq.xbar}
\bar{x}_{kernel} =\frac{1}{9}\sum_{x=1}^{3}\sum_{y=1}^{3}\left (
\hat{f}_{kernel}(x_{i,j}) \right).
\end{equation}
\begin{equation}\label{Eq.LSBfilter}
\hat{f}_{kernel}(x_{i,j}) =\left\lfloor \frac{f_{kernel}(x_{i,j})}{8} \right\rfloor\times8.
\end{equation}

In Eq.~\ref{Eq.LSBfilter}, $f_{kernel}$ shows the amount of gray-level of pixels in the 3$\times$3 kernel and in Eq.~\ref{Eq.xbar}, $\bar{x}_{kernel}$ shows the average of them. In order to deal with any possible problem in extraction phase, only 5 MSB-bits per pixel ${f}_{kernel}(x_{i,j})$ is used to calculate SD. Accordingly, in order to delete the three less importance bits of each pixel of kernel, the floor operator is used.
As mentioned in sub-step of decimal coding, for each 63-bits string an integer decimal number for the length of string and an integer decimal number for the decimal representation of string are calculated, which both of the mentioned numbers are placed in a range of 0-63 (6 bit for each integer decimal number). Therefore, 12 bits are needed to code a string with a length of 63-bit. The embedding algorithm places the related 12-bit in three LSB bits of four embeddable pixels. In order to increase the security, each produced 12-bit string were exclusive-ORed with a randomly produced 12-bit string by a common key between sender and receiver. 

\subsection{Extraction Step}

In this step, first, the message-contained embeddable pixels should be found using Eq.~\ref{Eq.stdDev}-\ref{Eq.LSBfilter}. Every 4 embeddable pixels contain 12-bit information related to a 63-bit length string. By extracting and concatenating all of them, a bit vector including information of document dimension (two 20-bit number), coordinates of rectangles (four 12-bit value for each block), and blocks content are obtained. With the location coordinates of each rectangular block, the information content of each block is easily placed in its place and the image of the complemented document is obtained. Then, in order to achieve a halftone image, it should be performed a complement operation. The different methods have been proposed to calculate the inverse halftone image of the gray-level image up to now. In the present research, the halftone image is convolved by a 3$\times$3 Gaussian kernel with a standard deviation of 0.5.

\section{Experimental Results}\label{lbl.experimentalresult}
In the present study, 20 standard gray-level image of USC-SIPI image database was used to evaluate the proposed method. Besides, the scanned document of an article~\cite{ref18} was used as a message (each page was scanned in the scale of 1774$\times$1288).

\begin{table*}[!htbp]
,\caption{Results of PSNR, SSIM and embedding rate for 20 standard images with different standard deviations and minimum rectangle size $4\times4$ }\label{tbl.standardresults}
  \centering
\begin{tabular}{l c c c| c c c| c c c}
&  \multicolumn{3}{c|}{Standard deviation (0)} &  \multicolumn{3}{c|}{Standard deviation (2.5)} &  \multicolumn{3}{c}{Standard deviation (5)}\\
\hline

Image & \rotatebox[origin=c]{90}{PSNR} & \rotatebox[origin=c]{90}{SSIM}& \rotatebox[origin=c]{90}{Embedding rate}  & \rotatebox[origin=c]{90}{PSNR} & \rotatebox[origin=c]{90}{SSIM}& \rotatebox[origin=c]{90}{Embedding rate} & \rotatebox[origin=c]{90}{PSNR} & \rotatebox[origin=c]{90}{SSIM}& \rotatebox[origin=c]{90}{Embedding rate}\\
\hline

Aerial & 37.36 & 0.95 & 9.91 & 37.66 & 0.96 & 6.97 & 38.99 & 0.98 & 6.08\\

Airplan & 37.11 & 0.87 & 9.91 & 38.49 & 0.91 & 6.80& 46.39 & 0.99 & 2.62\\

APC & 37.32 & 0.91 & 9.91 & 37.53 & 0.91 & 7.19 & 40.81 & 0.96 & 5.81\\

Barbara & 37.36 & 0.94 & 9.91 & 37.59 & 0.94 & 7.37 & 39.52 & 0.98 & 5.41\\

Boat & 37.33 & 0.94 & 9.91 & 37.46 & 0.94 & 9.62 & 39.15 & 0.97 & 5.36\\

Cameraman & 37.35 & 0.90 & 9.91 & 38.84 & 0.94 & 5.26 & 42.40 & 0.99 & 3.61\\

Couple &37.33 & 0.93 & 9.91 & 37.63 & 0.94 & 6.5 & 39.95 & 0.97 & 3.63\\

Elaine & 37.33 & 0.93 & 9.91 & 37.43 & 0.93 & 9.76 & 38.66 & 0.95 & 7.51\\
Goldhill & 37.31 & 0.93 & 9.91 & 37.63 & 0.94 & 7.07 & 39.30 & 0.97 & 5.68\\

House & 37.64 & 0.91 & 9.91 & 39.61 & 0.95 & 4.10 & 43.86 & 0.99 & 2.71\\

Jetplane & 37.36 & 0.91 & 9.91 & 38.14& 0.93 & 7.51 & 41.88 & 0.98 & 4.27\\

Lake & 37.32 & 0.93 & 9.91 & 37.51 & 0.94 & 9.46 & 39.35 & 0.97 & 5.61\\

Lena & 37.35 & 0.91 & 9.91 & 37.71 & 0.92 & 6.35 & 41.08 & 0.98 & 6.13\\

Man & 37.40 & 0.93 & 9.91 & 37.81 & 0.95 & 6.35 & 39.14 & 0.97 & 5.78\\

Pepper & 37.34 & 0.92 & 9.91 & 37.48 & 0.92 & 9.41 & 40.55 & 0.97 & 5.98\\

Truck & 37.32 & 0.93 & 9.91 & 37.50 & 0.93 & 9.29 & 39.39  & 0.96 & 6.23\\

Walk Bridge & 37.66 & 0.97 & 9.91 & 37.80 & 0.97 & 9.71 & 38.43 & 0.98 & 7.02\\

Blonde & 37.33 & 0.92 & 9.91 & 37.48 & 0.93 & 9.46 & 39.89 & 0.97 & 3.88\\

Dark Hair & 37.35 & 0.89 & 9.91 & 38.37 & 0.92 & 7.61 & 43.41 & 0.98 & 3.73\\

Zelda & 37.35 & 0.91 & 9.91 & 37.60 & 0.91 & 7.19 & 41.93 & 0.98 & 4.10\\
\hline
\textbf{Average} & \textbf{37.37} & \textbf{0.92} & \textbf{9.91} & \textbf{37.83} & \textbf{0.93} & \textbf{7.98} & \textbf{40.67} & \textbf{0.97} & \textbf{5.04}\\
\hline

\end{tabular}
\end{table*}

\begin{table*}[!htbp]
,\caption{Comparison of the PSNR and embedding rate of proposed method and methods~\cite{ref3} and~\cite{ref1}} \label{tbl.ref3-ref1}
  \centering
  \begin{tabular}{l| c c|  c c |  c c}
   & \multicolumn{2}{c|}{Proposed method} &  \multicolumn{2}{c}{Method~\cite{ref3}} & \multicolumn{2}{c}{Method~\cite{ref1}}\\
  \hline
  
  \rotatebox[origin=c]{90}{Image} & \rotatebox[origin=c]{90}{PSNR} & \rotatebox[origin=c]{90}{Embedding rate}& \rotatebox[origin=c]{90}{PSNR} & \rotatebox[origin=c]{90}{Embedding rate}& \rotatebox[origin=c]{90}{PSNR} & \rotatebox[origin=c]{90}{Embedding rate}\\
  \hline

  Jetplane & 38.14& 7.51 & 30.66 & 2.34 & 34.70 & 1.06   \\
  Lena     & 37.71& 6.35 & 31.94 & 2.34& 35.00 & 1.10   \\
  Pepper   & 37.48& 9.41 & 30.42 & 2.33& N/A & N/A  \\
  Lake     & 37.51& 9.46 & 28.99 & 2.41 &33.00 & 1.22   \\
  Man      & 37.81& 6.35 & 28.63 & 2.43 & N/A&  N/A  \\
  Barbara & 37.46& 9.62 &  N/A& N/A & 33.00 & 1.37  \\
    
  \hline
  \textbf{Average} & \textbf{37.68} & \textbf{7.61} & \textbf{30.12} & \textbf{2.37}& \textbf{33.92} & \textbf{1.18}
  \\
  \hline
  
  \end{tabular}
\end{table*}


\begin{table*}[!htbp]
,\caption{Comparison of the PSNR and embedding rate of proposed method and methods~\cite{ref4} and~\cite{ref6}}\label{tbl.ref4-ref6}
  \centering
\begin{tabular}{l| c c|  c c |  c c}
 & \multicolumn{2}{c|}{Proposed method} &  \multicolumn{2}{c}{Method~\cite{ref4}} &  \multicolumn{2}{c}{Method~\cite{ref6}} \\
\hline

\rotatebox[origin=c]{90}{Image} & \rotatebox[origin=c]{90}{PSNR} & \rotatebox[origin=c]{90}{Embedding rate}& \rotatebox[origin=c]{90}{PSNR} & \rotatebox[origin=c]{90}{Embedding rate}& \rotatebox[origin=c]{90}{PSNR} & \rotatebox[origin=c]{90}{Embedding rate} \\
\hline

Pepper      & 37.48 & 9.41 & 34.93& 3.95 & N/A& N/A \\
Lena     & 37.71& 6.35 & N/A & N/A & 37.97& 2.02   \\
Aerial      & 37.66 & 6.97 & N/A& N/A & 37.96& 2.02  \\
Jetplane      & 38.14 & 7.51 & 34.67 & 3.95 & 37.96& 2.02  \\
\hline
\textbf{Average} & \textbf{37.74} & \textbf{7.56} & \textbf{34.8} & \textbf{3.95}& \textbf{37.96} & \textbf{2.02}
\\
\hline
\end{tabular}
\end{table*}



\begin{table*}[!htbp]
,\caption{Comparison of the PSNR and embedding rate of proposed method and methods~\cite{ref8} and~\cite{ref7}}\label{tbl.ref9-ref7}
  \centering
\begin{tabular}{l| c c|  c c |  c c}
 & \multicolumn{2}{c|}{Proposed method} &  \multicolumn{2}{c}{Method~\cite{ref8}} &  \multicolumn{2}{c}{Method~\cite{ref7}(Edge detection=Sobel)} \\
\hline

\rotatebox[origin=c]{90}{Image} & \rotatebox[origin=c]{90}{PSNR} & \rotatebox[origin=c]{90}{Embedding rate}& \rotatebox[origin=c]{90}{PSNR} & \rotatebox[origin=c]{90}{Embedding rate}& \rotatebox[origin=c]{90}{PSNR} & \rotatebox[origin=c]{90}{Embedding rate} \\
\hline

Blonde      & 37.48 & 9.46 & 35.23& 5.65 & 37.31 & 3.04 \\
Pepper     & 37.48& 9.41 & 36.19 & 5.65 & 37.27& 3.05   \\
Jetplane      & 38.14 & 7.51 & 36.84& 4.77 & 33.85& 3.91  \\
Boat      & 37.46 & 9.62 & 37.00 & 5.70 & 33.57& 3.91  \\
\hline
\textbf{Average} & \textbf{37.64} & \textbf{9.00} & \textbf{36.31} & \textbf{5.43}& \textbf{35.50} & \textbf{3.47}
\\
\hline
\end{tabular}
\end{table*}



\begin{table*}[!htbp]
,\caption{Comparison of the PSNR, SSIM and embedding rate of proposed method and method~\cite{ref5}}\label{tbl.ref5}
  \centering
\begin{tabular}{l| c c c|  c c c}
 & \multicolumn{3}{c|}{Proposed method} &  \multicolumn{3}{c}{Method~\cite{ref5}}  \\
\hline

\rotatebox[origin=c]{90}{Image} & \rotatebox[origin=c]{90}{PSNR} &
\rotatebox[origin=c]{90}{SSIM} & \rotatebox[origin=c]{90}{Embedding rate}& \rotatebox[origin=c]{90}{PSNR} &
\rotatebox[origin=c]{90}{SSIM} & \rotatebox[origin=c]{90}{Embedding rate} \\
\hline

Barbara      & 37.59 & 0.96  & 6.97 & 36.99& 0.88 & 3.80  \\
Boat         & 37.46 & 0.94  & 9.62 & 39.23& 0.88 & 3.68  \\
House        & 39.61 & 0.95  & 4.10 & 40.23& 0.91 & 3.67  \\
Jetplane     & 38.14 & 0.93  & 7.51 & 40.90& 0.90 & 3.60  \\
Pepper       & 37.48 & 0.92  & 9.41 & 40.15& 0.93 & 3.66  \\
Truck        & 37.50 & 0.93  & 9.29 & 41.58& 0.95 & 3.61  \\
Zelda        & 37.60 & 0.91  & 7.19 & 41.79& 0.95 & 3.56  \\
Lena         & 37.71 & 0.92  & 6.35 & 40.20& 0.91 & 3.60 \\

\hline
\textbf{Average} & \textbf{37.88} & \textbf{0.93} & \textbf{7.55} & \textbf{39.88} & \textbf{0.91} & \textbf{3.64}\\
\hline
\end{tabular}
\end{table*}


\begin{table*}[!htbp]
,\caption{Comparison of the PSNR and embedding rate of proposed method and related works}\label{tbl.reportedresults}
  \centering
\begin{tabular}{l c c }
 Method & PSNR & Embedding rate (bpp) \\
\hline

 \cite{ref1}     & 33.92& 1.18 \\
  \cite{ref6}      & 37.96& 2.02\\
  \cite{ref3}      & 30.12& 2.37 \\
  \cite{ref7}      & 35.50& 3.47 \\
   \cite{ref4}      & 34.8& 3.95 \\
  \cite{ref8}      & 36.31& 5.43 \\
 \textbf{Proposed method}   & \textbf{37.83}& \textbf{7.98} \\
 \hline

\end{tabular}
\end{table*}

	\begin{figure}[h]
	 \centering
	 \includegraphics[width=0.9\columnwidth]{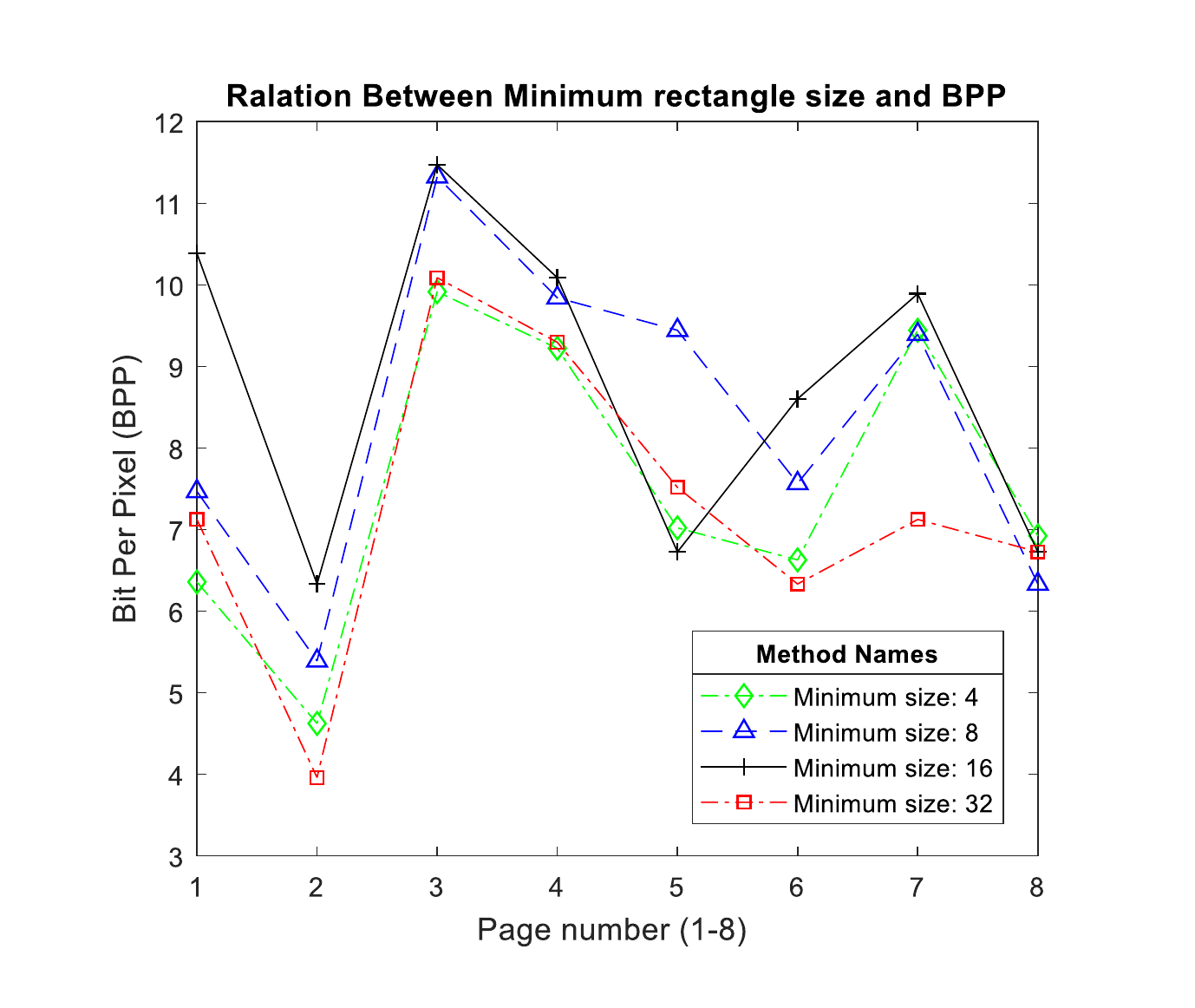}
	 \caption{Diagram of the information extraction and the embedding phase.}\label{fig.MeanAllPages}
	\end{figure} 
	
		\begin{figure}[h]
		 \centering
		 \includegraphics[width=0.9\columnwidth]{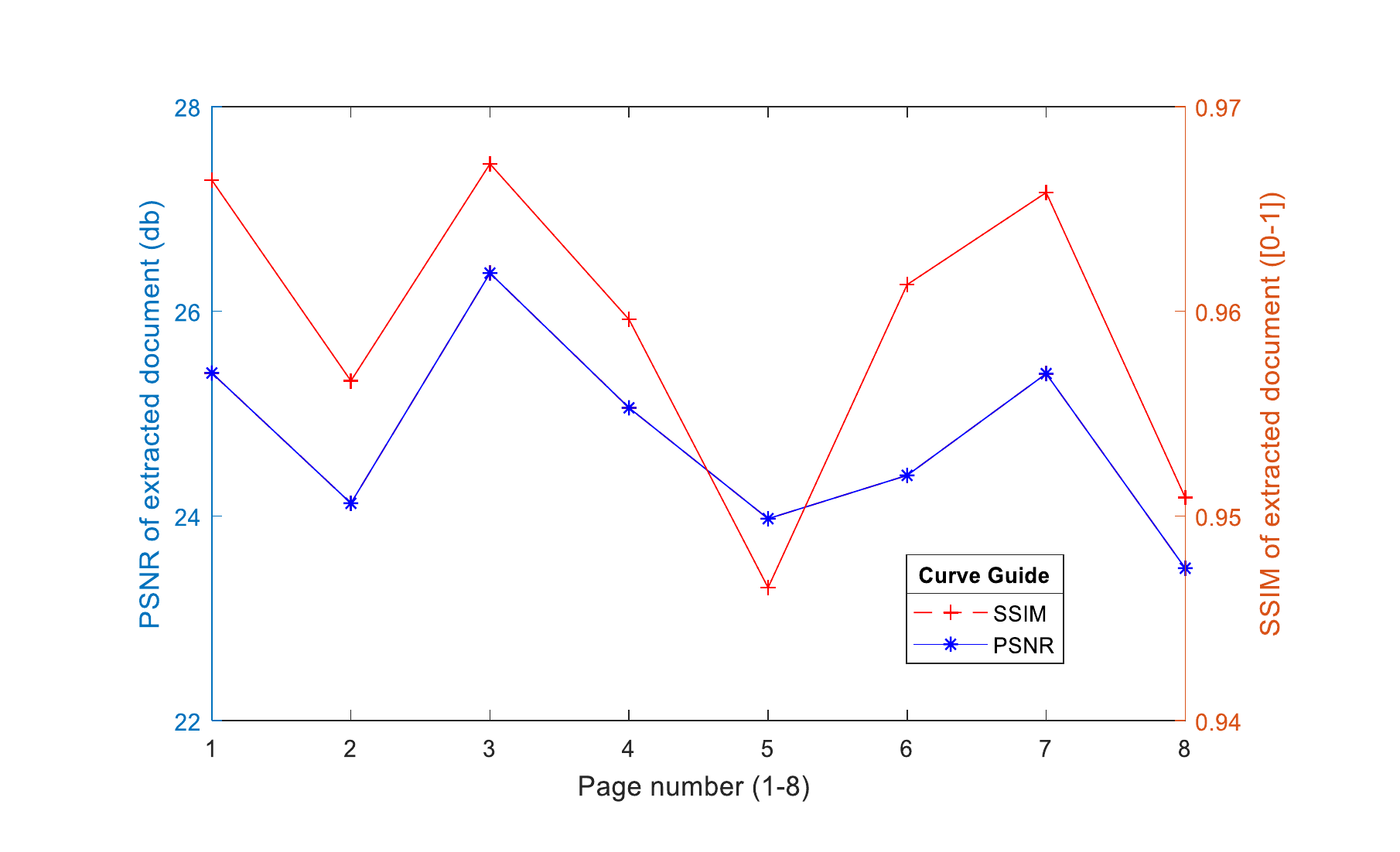}
		 \caption{Diagram of the information extraction and the embedding phase.}\label{fig.QualityOfextractedDocs}
		\end{figure}

The image quality criterion, which has been mentioned in many articles, is maximum signal to noise ratio or PSNR. The criterion shows amount the added noise to the image by embedding any watermark to it. Although this parameter is not exactly in accordance with the visual invisibility property of the watermark in the image, it provides a proper algebraic relationship for the optimal hiding of information in the image. The evaluation criterion is defined as Eq.~\ref{eq.psnr}:

\begin{equation}\label{eq.psnr}
  \textrm{PSNR}(f_c,f_w)=
  10\log_{10}\left [{\frac{max_{\forall(m,n)}f_{c}^2(m,n)}{\frac{1}{N_f}\sum_{\forall(m,n)}{\left( f_c(m,n)-f_{w}(m,n) \right)}^2}}\right].
\end{equation}

 The Eq.~\ref{eq.psnr} presents the maximum signal to noise ratio in terms of a decibel or dB. Where f is original image signal, $w$ is embedded a watermark, $f_{w}$ is watermarked image signal, ($m$,$n$) are indexes of the images, and $N_{f}$ is the number of pixels in images of $f_{c}$ and $f_{w}$. The more the criterion, the more ability of imperceptibility is. Often values of about 35 decibels are considered as acceptable values for this criterion. SSIM is another image quality criterion, which considers the structural property of the image. SSIM can be calculated by Eq.~\ref{eq.ssim}.

\begin{equation}\label{eq.ssim}
  \textrm{SSIM}(x,y)=(\frac{\sigma_{xy}}{\sigma_{x}\sigma_{y}}).(\frac{2\bar{x}\bar{y}}{(\bar{x})^2+(\bar{y})^2}).(\frac{2\sigma_{x}\sigma_{y}}{\sigma_{x}^2+\sigma_{y}^2}).
\end{equation}
The first term presents the correlation coefficient between two $x$ and $y$ images and shows the degree of linear dependency between the two images, which is in range of $[1, -1]$. The second term of Eq.~\ref{eq.ssim} that is in range of $[0,1]$ shows the average brightness of two images and when $x=y$ this term will be 1. The third term, which is in a range of $[0,1]$, is an estimation for color differentiation in x and y and the maximum value occurs when $\sigma_{x} =\sigma_{x}$. The SSIM and its updated versions is one of the important methods to determine the similarity of two images. In the present study, both mentioned criteria were used to evaluate the quality of embedded image and extracted document image. 

In Table~\ref{tbl.standardresults}, the results of applying the proposed method to the 20 selected standard image are shown. In this test, the minimum size of rectangles was 4$\times$4 and three different experimental thresholds such as 0, 2.5, and 5 were chosen for standard deviation to select the embeddable pixels. As shown in Table~\ref{tbl.standardresults}, for standard deviation 2.5, the embedding rate, PSNR, and SSIM criteria were 7.98 bpp, 37.83 dB and 0.93 in average state, respectively, which the value of embedding rate was significant and value of PSNR and SSIM were acceptable. 

In Tables~\ref{tbl.ref3-ref1}-\ref{tbl.ref5}, the comparison of the proposed method with similar works are presented. In this comparisons the same host images are used. Table~\ref{tbl.reportedresults} shows the summary of the average quality and the embedding rate of the proposed method in comparison with the related works. As seen, the proposed method has acceptable image quality and higher embedding rate. In Fig.~\ref{fig.MeanAllPages}, the proposed method is applied to eight different pages of an article~\cite{ref18}, as a message, and the Lena image is used as the host image. The mean embedding capacity of the proposed method for four different thresholds with the rectangular size of 4$\times$4, 8$\times$8, 16$\times$16, and 32$\times$32 were 7.51, 8.34, 8.77, and 7.27, respectively. The embedding capacity for the size of 16$\times$16 was more than three other forms. Fig.~\ref{fig.QualityOfextractedDocs} shows the quality of extracted document in terms of PSNR and SSIM for 8 different pages of an article. The mean of these values was 24.77 dB and 0.96 for all pages.

\section{Conclisions}\label{lbl.conclusions}
In the present research, a high capacity steganography method was introduced to embed the scanned documents as a message. The main ideas of the proposed method includes a) converting scanned document with gray-level to binary halftone image; b) proposing an algorithm to separate the content of the document from its background; c) compressing the message by ignoring the 0-bit stings behind the 1-bit value in process of converting binary bit string to its equivalent decimal value. The outputs resulting from conducting the proposed method, report the embedding capacity of 7.89 bpp and embedded image quality of 37.83 dB, which is higher and more effective than other methods with a random message. Therefore, the proposed method can be used for imperceptible transmission of text documents in the gray-level images.

One disadvantage of the proposed method is the ineffectiveness of calculating the halftone image and inverse of the halftone image. Design of a method to calculate the halftone and inverse halftone image, which is special for a text document, can improve the quality of the extracted document. Therefore, as a future work, it is suggested to design a method to calculate the halftone and inverse halftone image for the text documents.

As an another future work, it is suggested to use sparse representation and dictionary learning methods in order to more compress the text document images.

\section*{References}
\bibliographystyle{elsarticle-num}

\begin{thebibliography}{99}
\bibitem{ref001} Cox I, Miller M, Bloom J, Fridrich J, Kalker T. Digital watermarking and steganography. Morgan Kaufmann; 2007 Nov 23.\\

\bibitem{ref002} Soleymani SH, Taherinia AH. Double expanding robust image watermarking based on Spread Spectrum technique and BCH coding. Multimedia Tools and Applications. 2017 Feb 1;76(3):3485-503.\\

\bibitem{ref003}Soleymani SH, Taherinia AH. Robust image watermarking based on ICA-DCT and noise augmentation technique. InComputer and Knowledge Engineering (ICCKE), 2015 5th International Conference on 2015 Oct 29 (pp. 18-23). IEEE.\\

\bibitem{ref004}Haghighi BB, Taherinia AH, Harati A. TRLH: Fragile and blind dual watermarking for image tamper detection and self-recovery based on lifting wavelet transform and halftoning technique. Journal of Visual Communication and Image Representation. 2017 Oct 14.\\

\bibitem{ref005}Shi H, Li MC, Guo C, Tan R. A region-adaptive semi-fragile dual watermarking scheme. Multimedia Tools and Applications. 2016 Jan 1;75(1):465-95.\\

\bibitem{ref006}Qazanfari K, Safabakhsh R. A new steganography method which preserves histogram: Generalization of LSB++. Information Sciences. 2014 Sep 1;277:90-101.\\

\bibitem{ref007}Jung KH. A high-capacity reversible data hiding scheme based on sorting and prediction in digital images. Multimedia Tools and Applications. 2017 Jun 1;76(11):13127-37.\\

\bibitem{ref008}Qian Z, Zhang X. Reversible data hiding in encrypted images with distributed source encoding. IEEE Transactions on Circuits and Systems for Video Technology. 2016 Apr;26(4):636-46.\\

\bibitem{ref009}Liao X, Shu C. Reversible data hiding in encrypted images based on absolute mean difference of multiple neighboring pixels. Journal of Visual Communication and Image Representation. 2015 Apr 1;28:21-7.\\

\bibitem{ref010}Zhang X, Long J, Wang Z, Cheng H. Lossless and reversible data hiding in encrypted images with public-key cryptography. IEEE transactions on circuits and systems for video technology. 2016 Sep;26(9):1622-31.\\

\bibitem{ref011}Li X, Li B, Yang B, Zeng T. General framework to histogram-shifting-based reversible data hiding. IEEE Transactions on Image Processing. 2013 Jun;22(6):2181-91.\\

\bibitem{ref012}Wang ZH, Lee CF, Chang CY. Histogram-shifting-imitated reversible data hiding. Journal of systems and software. 2013 Feb 1;86(2):315-23.\\

\bibitem{ref013}Chen B, Wornell GW. Quantization index modulation methods for digital watermarking and information embedding of multimedia. Journal of VLSI signal processing systems for signal, image and video technology. 2001 Feb 1;27(1-2):7-33.\\

\bibitem{ref1}Tang M, Zeng S, Chen X, Hu J, Du Y. An adaptive image steganography using AMBTC compression and interpolation technique. Optik-International Journal for Light and Electron Optics. 2016 Jan 1;127(1):471-7.\\

\bibitem{ref2}Tang M, Hu J, Song W. A high capacity image steganography using multi-layer embedding. Optik-International Journal for Light and Electron Optics. 2014 Aug 1;125(15):3972-6.\\

\bibitem{ref3}Jung KH, Yoo KY. High-capacity index based data hiding method. Multimedia Tools and Applications. 2015 Mar 1;74(6):2179-93.\\

\bibitem{ref4}Jana B. High payload reversible data hiding scheme using weighted matrix. Optik-International Journal for Light and Electron Optics. 2016 Mar 1;127(6):3347-58.\\

\bibitem{ref5}Balasubramanian C, Selvakumar S, Geetha S. High payload image steganography with reduced distortion using octonary pixel pairing scheme. Multimedia tools and applications. 2014 Dec 1;73(3):2223-45.\\

\bibitem{ref6}Kanan HR, Nazeri B. A novel image steganography scheme with high embedding capacity and tunable visual image quality based on a genetic algorithm. Expert Systems with Applications. 2014 Oct 15;41(14):6123-30.\\

\bibitem{ref7}Bai J, Chang CC, Nguyen TS, Zhu C, Liu Y. A high payload steganographic algorithm based on edge detection. Displays. 2017 Jan 1;46:42-51.\\

\bibitem{ref8}Soleymani SH, Taherinia AH. High capacity image steganography on sparse message of scanned document image (SMSDI). Multimedia Tools and Applications. 2017 Oct 1;76(20):20847-67.\\

\bibitem{ref10}Liu L, Chen W, Zheng W, Geng W. Structure-aware error-diffusion approach using entropy-constrained threshold modulation. The Visual Computer. 2014 Oct 1;30(10):1145-56.\\

\bibitem{ref11}Li X. Edge-directed error diffusion halftoning. IEEE Signal Processing Letters. 2006 Nov;13(11):688-90.\\

\bibitem{ref12}Singh YK. Generalized error diffusion method for halftoning. InElectrical, Computer and Communication Technologies (ICECCT), 2015 IEEE International Conference on 2015 Mar 5 (pp. 1-6). IEEE.
\\

\bibitem{ref13}Zhou Z, Arce GR, Di Crescenzo G. Halftone visual cryptography. IEEE transactions on image processing. 2006 Aug;15(8):2441-53.\\

\bibitem{ref14}Alasseur C, Constantinides AG, Husson L. Colour quantisation through dithering techniques. InImage Processing, 2003. ICIP 2003. Proceedings. 2003 International Conference on 2003 Sep 14 (Vol. 1, pp. I-469). IEEE.\\

\bibitem{ref15}Chung KL, Wu ST. Inverse halftoning algorithm using edge-based lookup table approach. IEEE Transactions on Image Processing. 2005 Oct;14(10):1583-9.\\

\bibitem{ref16}Liu YF, Guo JM, Lee JD. Inverse halftoning based on the Bayesian theorem. IEEE Transactions on Image Processing. 2011 Apr;20(4):1077-84.\\

\bibitem{ref17}Finkel RA, Bentley JL. Quad trees a data structure for retrieval on composite keys. Acta informatica. 1974 Mar 1;4(1):1-9.\\

\bibitem{ref18}Hien TD, Nakao Z, Chen YW. Robust multi-logo watermarking by RDWT and ICA. Signal Processing. 2006 Oct 1;86(10):2981-93.\\

\bibitem{ref19}"SIPI Image Database". Sipi.usc.edu. N.p., 2016. Web. 25 Mar. 2016.

\end{thebibliography}

\end{document}